# Leveraging Machine Learning Approaches to Predict the Impact of Construction Projects on Urban Quality of Life

## Zhengbo Zou[1] and Semiha Ergan[2]

**ABSTRACT**

According to the World Bank, more than half of the world's population now lives in cities, creating burdens on degraded city infrastructures and driving up demands for new ones. Construction sites are abundant in already dense cities and have unavoidable impacts on surrounding environments and residents. However, such impacts were rarely quantified and made available to construction teams and local agencies to inform their planning decisions. A challenge in achieving this was the lack of available data that can provide insights about how urban residents respond to changes in their environment due to construction projects. Wider availability of data from city agencies nowadays provides opportunities for such analysis. This study provides the details of a machine learning based approach that enables the prediction of impact of construction projects on quality of life in urban settings through the quantification of changes on quality of life indicators (e.g., noise, air quality, traffic) in cities, inferred by open city data. This study also details the evaluation of the approach through a case study, which uses data from publicly available construction projects information and open data portal in New York City. Historical 311 Service Requests from New York City along with twenty-seven road reconstruction projects were used as testbeds. The predicted results showed that the construction projects experienced complaints such as "noise", "air quality", and "sewer" at the beginning of construction, and "sanitation" and "waste" towards the end of construction. The approach is capable of providing insights for government agencies and construction companies to take proactive actions based on expected complaint types/counts through different phases of construction, and avoid negative influences ahead of time.

---

[1] PhD Candidate, Department of Civil and Urban Engineering, New York University, 15 MetroTech Center, Brooklyn, NY, 11201; email: zz1658@nyu.edu

[2] Assistant Professor, Department of Civil and Urban Engineering, New York University, 15 MetroTech Center, Brooklyn, NY, 11201; email: semiha@nyu.edu

## 1. INTRODUCTION

Construction sites are common scenes in already dense cities. In NYC alone, on average, 20 permits per day were issued for new construction in 2016 (DOB 2016). Construction projects have indisputable impacts on various aspects of city life, such as environmental effect (i.e., noise pollution, air pollution, and construction waste), public safety (i.e., fatality of general public, emergency response time), and traffic (i.e., traffic delays, lane closing). Previous research that looked at the impact of construction on the general public mostly focused on one specific area of interest (i.e., noise, air quality, and public safety), instead of providing a holistic and quantitative view of construction projects' impact on nearby residents, let alone enabling predictive analysis (Butera et al. 2015; Hammad et al. 2016; Kumar et al. 2015). As a result, government agencies and construction companies only react to complaints reported from residents close to construction sites, instead of proactively preventing the negative influences on residents' quality of life that are expected to occur (Zou and Ergan 2018).

Quality of Life (QoL) is defined as the measurement of satisfaction and happiness in life with respect to where we live (Days 1987). Improving urban QoL has been one of the main goals for policy makers and urban researchers. Governments and International Organization of Standardization (ISO) have developed indicators for measuring QoL to monitor residents' satisfaction of life, and benchmark performance of cities around the globe. QoL indicators are separated into two groups as, subjective indicators (e.g., self-perceived health condition, self-perceived happiness level), which measure satisfaction of life at individual levels; and objective indicators (e.g., physical environment, public safety, education), which measure the environmental and societal aspects of QoL (Seik 2000). Previously, QoL indicators have been studied via public surveys issued by the census bureau, researchers, or local agencies aiming at collecting data for certain indicators such as income and education levels (US Census Bureau 2016, NYC CBC 2017). With the wider availability of open city data and machine learning techniques, an increasing number of researchers and practitioners started to use open city data as a social database to examine urban problems regarding residents' QoL (Barbosa et al. 2014). This open data plus machine learning methodology has already proven to be valuable in various empirical implementations, such as predicting air quality levels and crime rates (Liu et al. 2015; Zhan et al. 2017), making sense of taxi data for understanding traffic conditions (Ferreira et al. 2013; Smith et al. 2017; Zou and Sha 2018), and  monitoring urban noise problems (Segura-Garcia et al. 2015).

Although machine learning based methods in urban settings have been implemented to solve a variety of urban problems, little effort has been put into understanding the impact of construction projects on QoL in urban settings using open city data. The challenge of doing so is two-folded. First, there is a lack of data regarding construction projects. Construction companies hold their historical data (i.e., project estimate, schedules, and bidding contracts) as confidential materials and use it for competitive advantage in markets (Egbu 2004). Luckily, nowadays in public projects, generic construction data (e.g., project start/end times, milestones, total bid price, etc.) has been given public access in some cases, such as NYC open data regarding road constructions. Another challenge is the lack of knowledge on how construction caused urban issues map to the generic QoL indicators. Construction activities affect a wide variety of QoL indicators (i.e., noise, air quality, public safety), and the relationship between a specific construction phase and its impact on quality of life indicators is still unclear.

The objective of this study is to provide a new quantitative understanding of the impact of construction projects on urban QoL through a machine learning based approach. Specifically, this study aims to predict how much and what type of an impact a new construction project, with similar characteristics (e.g., location, project type) to the previously analyzed ones, would have on residents' QoL. The purpose is to inform construction planners and city officials about proactive measures for minimizing the negative impacts. The approach builds on the mapping of construction caused urban issues to the QoL indicators, and is composed of two steps, as: (1) filtering and feature selection of available datasets related to construction projects and urban QoL, and (2) development of machine learning models to "learn" from the past construction projects and predict the impact of new projects on QoL. For implementation and evaluation of the approach, road reconstruction projects in New York City were selected as testbeds, as horizontal projects (e.g., road construction) have a wider footprint to impact existing infrastructure and residents. Open datasets used in the evaluation of the approach were 311 service requests (which document non-emergency municipal service requests, including complaints received related to construction activities) and Department of Design and Construction (DDC) construction projects database. For this specific implementation, the prediction models aimed to predict (1) the number of complaints received per complaint type, and (2) the complaints ratio, which is defined as the number of complaints per complaint type divided by the total number of complaints received in the analysis period.

This paper contributes to the existing body of knowledge on construction and QoL research in the following ways: (1) development of a novel machine learning based approach to help construction teams to quantify construction projects' impact on urban QoL by leveraging city datasets; and (2) providing empirical evidence on construction projects' impact on QoL in urban settings by conducting case studies using data from construction projects in New York City.

## 2. BACKGROUND

This study builds on and extends the work on (a) defining construction related factors on Quality of Life (QoL), and (b) standardization efforts and previous research on defining and measuring QoL. The point of departure for this research is at the intersection of the research conducted in these two areas, where factors defined as influential on QoL in the construction literature are mapped to the indicators defined in standards/previous research.

### 2.1. Research on Defining Construction Related Factors on QoL

When the literature in the construction domain is analyzed, it is apparent that previous research mostly focused on analyzing the impact of individual urban issues caused by construction projects (e.g., heavy hauling and impact on traffic), instead of a holistic understanding of how construction sites impact the QoL of residents (Camagni et al. 2002; Hammad et al. 2016; Ivaskova et al. 2015). Major construction related factors that create public discomfort can be grouped under five categories as noise, air pollution, waste, public safety issues, and traffic issues.

Construction noise is one of the main contributors of urban acoustic pollution, affecting citizens' health and causing discomfort (Ballesteros et al. 2010), including psychological (e.g., annoyance, stress) and physical (e.g., hearing loss, cardiovascular deceases) discomfort (Hammad et al. 2016; Lee et al. 2015). City agencies often require the monitoring of noise levels at construction sites and keeping the noise below certain thresholds (e.g., for the use of vibrating pile driver, the limit is 101 dB in NYC) (NYC Environmental Protection 2007). Aside from the government regulations, researchers have made strides in construction noise monitoring and mitigation with the goal of helping construction companies to avoid shutdowns and reduce cost (Zou et al. 2007). Construction noise monitoring focuses on the placement of acoustic sensors to accurately record the decibel level of the heavy construction machine and overall construction sites (Shen et al. 2005). On the other hand, construction noise minimization includes two distinct directions of research, site layout optimization and construction schedule optimization. Site layout optimization aims to place site material, machinery and temporary structures strategically to occlude noise. Schedule optimization tries to minimize construction noise by separating noisy operations on construction sites to avoid combined acoustic pollution caused by multiple construction activities (Hammad et al. 2016).

Atmospheric emission is another key polluting factor that is associated with construction activities. Environmental audits, which document the energy used and air pollution produced during constructions, show that air emissions (e.g., fine particulate matter (PM2.5 and PM10) and $CO_2, SO_2, NO_x$) are byproducts from construction activities and should be monitored (Cole and Rousseau 1992; Kampa and Castanas 2008). Excessive respirable dust generated during construction (e.g., excavation, interior finishing) can also lead to air pollution and eventually health concerns (Lumens and Spee 2001). Previous research on air pollution caused by construction focused on two aspects: pollution generated during the process of material manufacturing (i.e., steel, aluminum, and cement), and pollution caused during the construction phase (i.e., material transportation) (Chan and Yao 2008; Cole and Rousseau 1992). Studies on finding renewable construction materials and clean energy sources have created opportunities to reduce air pollution in the process of manufacturing construction material (Cheng and Hu 2010; Cho et al. 2010; Herrmann et al. 1998). During the construction phase, material transportation routes and schedule optimization aim to reduce the cost of transportation as well as the environmental effects caused by heavy hauling (Zhou et al. 2010). Dust control approaches, such as reducing on-site manufacturing and proper protection during the demolition process, were studied to alleviate the dust issue on construction sites (Tjoe Nij et al. 2003; Wu et al. 2016).

Waste generated during construction projects posts serious problems for construction companies and government agencies for both economic and environmental reasons. Economically, the cost of dumping construction waste has been increasing for the past few decades, raising cost for construction companies (Rao et al. 2007). On the other hand, construction waste dumped at landfills causes environmental concerns (Poon et al. 2004). Studies to address this concern are divided into two parts, including waste management planning in the construction development phase, and recycling/reusing of the demolished construction materials. Simulation studies were often used in the project planning phase to predict construction waste sources and provide optimized site layout and disposal plans to minimize waste generated onsite (Wang et al. 2014; Yuan 2013). For recycling construction waste, researchers proposed physical tracking devices for reusable construction parts, and created recycle strategies for construction sites (Li et al. 2005; Shen et al. 2004).

Public safety is another crucial aspect of urban life influenced by construction activities. Ireland Health and Safety Authority reported almost one fatality per month from the general public due to construction activities (Suraji et al. 2001). However, research efforts on construction safety studies mostly focused on construction worker safety and safety regulations (Elbeltagi et al. 2004; Tam et al. 2002, 2004). Construction impact on public safety is mainly studied in projects where interactions with the public is unavoidable, such as airport reconstruction and hospital renovations (Toor and Ogunlana 2010). Available studies are also case examinations that reflect on immediate tragedies of public fatalities caused by construction activities (Müngen and Gürcanli 2005; Wang et al. 2008). One of the reasons for the lack of research in the public safety domain is the lack of data. While agencies such as Occupational Safety and Health Administration (OSHA) keeps track of worker fatalities, the raw data for analyzing construction caused public safety problem is limited.

Infrastructure construction projects also cause traffic delays because of the extended construction sites and road closures. Previous research on minimizing the traffic influences caused by construction activities considered traffic slowdowns as part of the construction cost, and then apply optimization methods to reduce the overall cost as much as possible (Lee 2009). Another factor being considered for construction affected traffic situation is the construction schedule. Optimization methods have been applied to shorten the construction duration, hence reducing the total amount of time wasted in traffic due to construction projects (Carr 2000; Lee et al. 2005; Yepes et al. 2015).

## 2.2. Standardization Efforts and Previous Research on Defining and Measuring QoL

The efforts under this subtitle fall into two categories as studies/efforts to define metrics to quantify QoL systematically, and studies that used such metrics for quantifying aspects of QoL in selected regions. Researchers and standardization organizations have been working on defining indicators of QoL. A landmark study in the early 2000s (Seik 2000) included 18 QoL categories as: social life, working life (career), family life, education, wealth, health, religion, leisure, self-development, housing, media, politics, consumer goods, public utilities, transportation, health care, environment, and public safety. Later survey studies often followed similar design paradigms for the survey categories (Das 2008; McMahon 2002; Santos and Martins 2007). The objective of these survey studies was to define a comprehensive list of subjective indicators of QoL and establish a mature QoL surveying and analyzing system, which can be used in future studies for consistent monitoring and comparison of citizens' subjective responses of QoL.

On the other hand, standardization bodies aim to define QoL metrics that are quantifiable. These metrics are objective indicators of QoL, which are mainly covered in the International Organization of Standardization (ISO) *Sustainable Develop of Communities – indicators for city services and quality of life*. There are 17 performance categories (i.e., objective indicators) defined for benchmarking across cities, including economy, education, energy, environment, finance, fire and emergency response, governance, health, recreation, safety, shelter, solid waste, telecommunication and innovation, transportation, urban planning, waste water, and water and sanitation (ISO 2014).

In recent years, the popularity of big data and machine learning techniques sparked new methodologies for objectively measuring QoL (Ferreira et al. 2013; Liu et al. 2015). Various objective QoL indicators (e.g., transportation, public health, and economics) were studied using vast amount of data collected privately or publicly (Ferris et al. 2010; Ruths and Pfeffer 2014; Zheng et al. 2013). In the transportation domain, taxi trips data in New York City has been analyzed using spatial-temporal algorithms to gain insights on taxi ridership and social events in the city (Doraiswamy et al. 2014; Ferreira et al. 2013; Phithakkitnukoon et al. 2010). For public health, researchers provided prediction tools for the spread of diseases by combining various data sources (e.g., Google searches, Tweets and hospital visiting records) (Lee et al. 2010; Santillana et al. 2015). For economics, administrative data from agencies such as Internal Revenue Service and Center of Medicare and Medicaid have been used to predict economic behavior (Bose and Mahapatra 2001; Einav and Levin 2014). However, no studies were focused on analyzing construction projects' impact on urban QoL indicators (Zou and Ergan 2018). This paper fills the gap of quantitative construction impact study on QoL in urban settings. In addition, as a result of the literature review, the authors mapped the construction related factors to QoL indicators, as shown in Figure 1. This mapping will be used as a given when building prediction models using the approach.

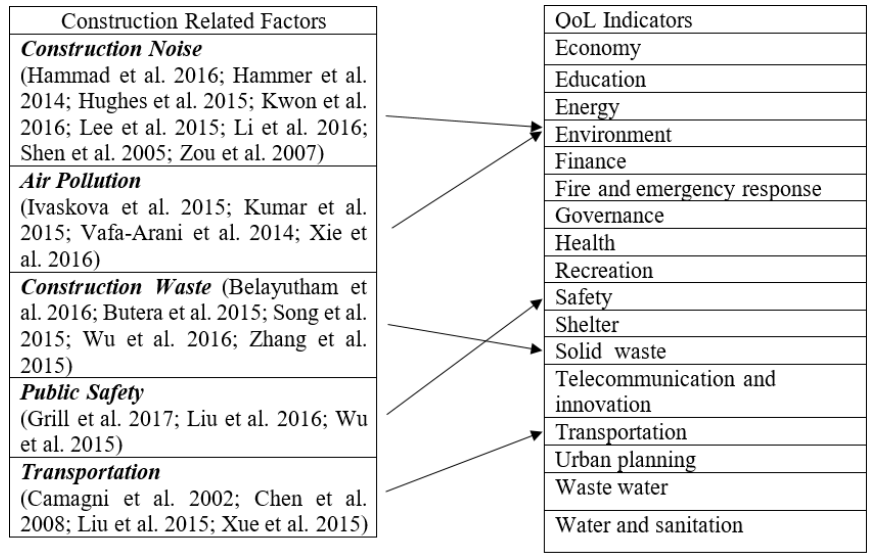

**Figure 1. Mapping between Construction Related Factors and QoL Indicators.**

## 3. METHODOLOGY

This study proposes a machine learning based approach for the quantification, prediction, and interpretation of impact of construction projects on urban QoL. In methodology, we introduce a generalized approach that leverages urban datasets and machine learning methods to quantify construction projects' impact on urban QoL, while in the evaluation section we validate the approach by implementing it on real data from New York City. In a nutshell, the proposed approach has two main steps. The first step includes pre-processing of datasets with (a) filtering of construction projects and open city data to create a subset of data for a selected analysis period, and (b) feature selection to eliminate the non-essential fields in datasets that are not statistically significant to the final prediction targets. The second step is building prediction models using the selected subset of features from the filtered datasets. Supervised learning methods are used here since the objective of the study is to predict the impact of construction projects on urban QoL, and such impacts can be defined mathematically as numerical measurements of direct or indirect QoL indicators (e.g., number of 311 complaints from residents about construction activities nearby). Finally, the best performing models are used as prediction models for new construction projects.

### 3.1. Step 1. Data Pre-Processing: Filtering and Feature Selection

In the pre-processing process, construction projects in the available project database should be filtered to create a needed subset. Filters such as the construction type (e.g., new building construction, road construction), construction location (e.g., dense cities, suburbs), construction commencement date, and construction duration should be defined ahead of time based on the characteristics of the new construction projects for which the impact analysis will be performed (Figure 2 first box). On the other hand, open city data regarding QoL indicators (e.g., 311 complaints received regarding construction) should also be filtered so that it only contains information related to the area of interest for the data analysis. For example, one can define the analysis period for construction projects to be two years (i.e., one year before the start date of projects and one year after their start dates), and data that is beyond this analysis period should be omitted.

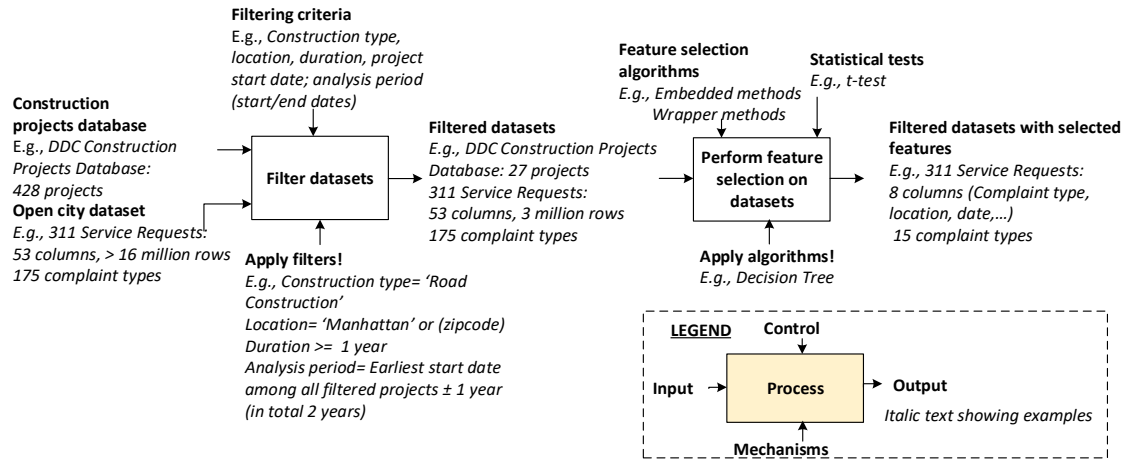

**Figure 2. IDEF0 Diagram Showing the Data Pre-Processing Step.**

Feature selection (Figure 2 second box) of open city data is crucial, since poorly selected features could negatively affect the accuracy of prediction models (Dash and Liu 1997). The significance level of each field from the selected data sources should be examined, and only the significant ones should to be retained for further analysis. In general, feature selection algorithms can be divided into two groups, filter and wrapper methods (Jain and Zongker 1997). As the name suggests, filter methods test each feature's significance level to a statistical model containing that feature, and eliminate it if the feature is not found to be statistically significant in the final prediction model. Different criteria could be used for the calculation of significance level, such as t-test and log-likelihood ratio test. On the other hand, wrapper methods do not directly apply statistical tests on the features. Alternatively, they build prediction models containing all subset configurations of the features, and select the most suitable ones by comparing the prediction models' performance. Although wrapper methods are often time consuming, they can exhaustively test all configurations of the features (Dash and Liu 1997). Therefore, in this study, a wrapper method (i.e., Decision Tree) was used to conduct feature selection during the evaluation phase.

## 3.2. Step 2. Building Prediction Models

In this step, prediction models are built using various machine learning algorithms. The input for this step is the pre-processed datasets from the previous step. The specific algorithm selection (i.e., supervised vs unsupervised) depends on the purpose of the study. If the objective is to predict an exact number or a class of a QoL indicator, supervised regression or classification methods are needed. On the other hand, if the objective is to explore behaviors of certain QoL indicators, unsupervised clustering techniques should be applied. An overview of this step with illustrative examples is provided in Figure 3.

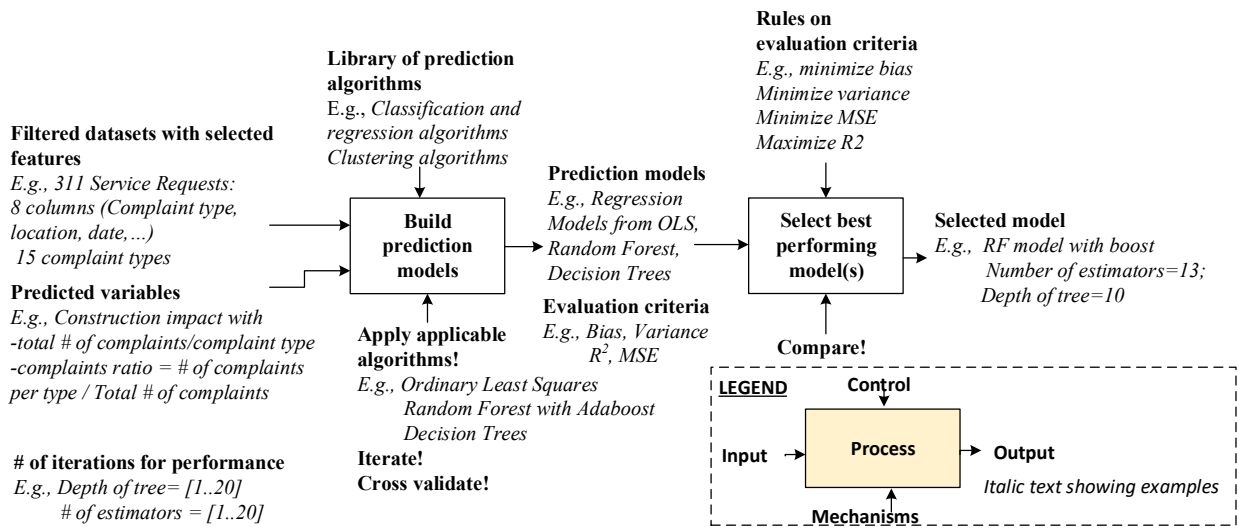

**Figure 3. IDEF0 Diagram Showing the Building Prediction Models.**

For this study, the objective is to predict the impact of construction projects on urban QoL. As will be discussed later in the paper, the impact is measured as numerical values. Such metrics are best captured with regression models, hence regression methods were used in this approach to build prediction models (Figure 3 first box). Specifically, Ordinary Least Squares (OLS), Random Forest (RF) with Adaboost, and Decision Tree (DT) were implemented when evaluating the approach. By including multiple machine learning algorithms, we evaluate model performances and suggest the best performing one during prediction. OLS was used as it is one of the simplest ways of implementing linear regression (Craven and Islam 2011), and is commonly used as a benchmarking model for other machine learning algorithms to compare to. Since the assumption of linearity is almost certainly too aggressive for urban data, linear regression was expected to produce the least accurate result for our problem.

Random Forest (RF) with Adaboost was also used in this approach because of its high accuracy as a regression model, and its capability of being used without the need for extensively large datasets or long training times. Earlier studies that compared ten most widely used supervised learning techniques on empirical datasets showed that Random Forest being the second-best performer, only after Boosted Trees with tuned parameters (Caruana and Niculescu-Mizil 2006). RF is one of the ensemble algorithms, which creates a "bag" of base models, and uses the aggregation of base model predictions as the final result. Adaboost was also implemented in the RF algorithm. Adaboost creates multiple base models in order, and improve its own performance based on the previous model's errors. In the RF algorithm, the base model is often selected as Decision Tree (DT). As a result, this study included DT as a comparison to examine if the addition of bagging and boosting created any improvement on the prediction results. Decision Tree is created based on a Tree Structure. Branches in regression trees are carrying weights that represent the confidence level of the choices made by the branches. The final regression value is a weighted aggregated value from all leaf nodes. As a last step, cross validation should be implemented for each algorithm to tune parameters of the models (Figure 3 first box). In this study, we implemented 10-fold cross validation.

The best performing prediction model can be selected using a series of metrics, such as bias, variance, R-Squared value, and Mean Squared Error (MSE), as shown in Figure 3 (second box). R-Squared value is commonly regarded as the first choice of benchmarking parameter to use (Salakhutdinov et al. 2007). Models with high R-Squared values have higher probability of correctly predicting unseen data. Bias and variance can also be calculated for models with different model parameters, with the goal of finding a low bias and low variance model. Low bias means less sum of error for the prediction model, whereas low variance means high confidence of the predicted value falls near the mean of true values. However, when bias and variance move on opposite directions with the change of model parameters, MSE can be used to moderate between model bias and variance. Because MSE is calculated as the sum of squared bias and variance (Salakhutdinov et al. 2007), minimizing MSE can lead to a balance where the model produces both low bias and variance. Once the final prediction model is selected based on model performances, predictions on new construction projects can be performed, and prediction results can be interpreted for future proactive actions.

## 4. EVALUATION OF THE APPROACH

This section provides the details of the implementation and evaluation of the approach using specific datasets (i.e., construction data for road reconstruction in Manhattan and 311 service requests regarding construction complaints). The implementation results are discussed for each step of the approach in the following subsections.

### 4.1. Overview of the Datasets Used for Evaluation

This study used open city data to measure construction projects' impact on urban QoL. The construction data was collected from the online project repository from New York City Department of Design and Construction (NYC DDC 2018). To measure the construction projects' impact, 311 Service Requests data was used. 311 data serves as a bridge to map urban QoL indicators to construction projects, since numerus types of 311 Service Requests are related to construction activities and show statistically significant differences before and during construction times. 311 Service Requests data was collected from New York City's online open data portal (DoITT 2018).

### 4.1.1. Road Reconstruction Projects in Manhattan

The New York City Department of Design and Construction (NYC DDC) serves as the project manager for the city's capital projects. NYC DDC's portfolio includes a wide variety of construction projects, such as horizontal projects (i.e., roads, bridges), vertical projects (i.e., buildings, towers), and renovation projects. In this study, construction projects containing road reconstruction were selected as testbeds due to their larger impact zones. Road reconstruction projects have the potential to affect many residents' QoL due to the extended sites, heavy machinery involved, and road closures, while the scale of influence of vertical construction projects is often limited to the construction block.

In order to create a precise prediction model for the road reconstruction projects, a series of constraints were applied to the NYC DDC project pool to select a subset. Firstly, the location of analysis was set as the borough of Manhattan, due to its large population and constant ongoing road constructions. Secondly, the analysis period of each construction project was set as two years, (i.e., one year before the construction, using as baseline, and one year after the construction commencement). After applying these constraints, 27 projects were selected from the NYC DDC pool of 428 capital projects in total (as seen in Table 1).

**Table 1. Summary of the Construction Projects Included in the Analysis.**

| # | Start | Duration | Zip | # | Start | Duration | Zip |
|---|---|---|---|---|---|---|---|
| 1 | 08/05/13 | 4 years | 10004 | 15 | 01/31/17 | 2 years 6 months | 10002 |
| 2 | 09/23/13 | 1 year 9 months | 10034 | 16 | 10/27/14 | 2 years | 10014 |
| 3 | 04/14/14 | 2 years 2 months | 10016 | 17 | 07/06/15 | 2 years | 10012 |
| 4 | 04/15/14 | 2 years 6 months | 10035 | 18 | 06/29/15 | 1 year | 10065 |
| 5 | 07/22/14 | 1 year 4 months | 10030 | 19 | 05/15/16 | 1 year | 10028 |
| 6 | 11/24/14 | 1 year 7 months | 10028 | 20 | 01/20/14 | 1 year 6 months | 10032 |
| 7 | 12/31/14 | 2 years | 10014 | 21 | 06/30/15 | 1 year | 10039 |
| 8 | 06/29/15 | 1 year | 10021 | 22 | 06/01/16 | 2 years | 10013 |
| 9 | 01/04/16 | 3 years | 10001 | 23 | 01/06/15 | 1 year 6 months | 10040 |
| 10 | 02/15/16 | 5 years | 10007 | 24 | 08/05/13 | 3 years | 10003 |
| 11 | 03/07/16 | 1 year | 10128 | 25 | 09/16/13 | 2 years | 10007 |
| 12 | 05/31/16 | 1 year 1 months | 10033 | 26 | 10/20/16 | 2 years | 10016 |
| 13 | 06/01/16 | 2 years | 10031 | 27 | 03/01/17 | 1 year 6 months | 10007 |
| 14 | 06/27/16 | 2 years 6 months | 10038 | | | | |

### 4.1.2. 311 Service Requests from 2010 to 2017

311 Service Requests data from 2010 to 2017 was used to quantify construction projects' impact on urban QoL, since a subset of complaints are construction related. The selection of construction related complaints will be introduced in the following sections. 311 requests document non-emergency complaints from NYC residents. The dataset includes 53 columns and more than 16 million rows, containing 175 different types of complaints. As a pre-processing step, repetitive information was excluded (e.g., *Intersection* and *Cross Streets*). Furthermore, any data with a time-stamp earlier than one year before the earliest start time of the analyzed construction projects was removed, because it falls out of the analysis period. Finally, to only retain the columns that are impactful for the prediction models, we implemented decision tree as a wrapper method for feature selection. In Decision Trees, the level of a node is decided by the weight that node carries. As shown in Figure 4, "complaint type" has the highest weight (root node), while "incident zip code" and "created date" (level one) having the second and third highest weights. Due to the space constraints, only part of the DT is shown in Figure 4. After feature selection, a snippet of the service requests data is provided in Table 2.

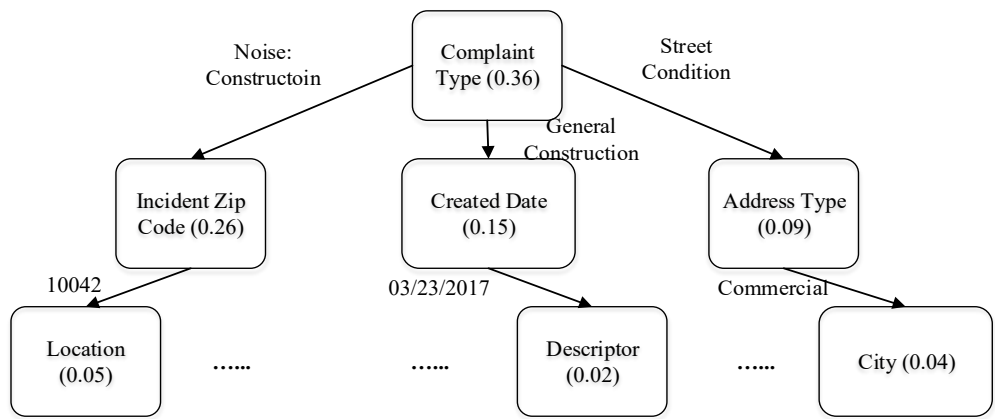

**Figure 4. Embedded Feature Selection using Decision Tree.**

**Table 2. A Snippet from 311 Service Request Data.**

| Unique Key | Created Date | Agency | Complaint Type | Descriptor | Incident Zip | Borough | … |
|---|---|---|---|---|---|---|---|
| 36154062 | 5/11/17 | DOB | General construction | Construction | 10002 | Manhattan | |
| 36154109 | 5/10/17 | DOT | Street condition | Construction caused congestion | 10075 | Manhattan | |

### 4.2. Introduction of the Target Variables

During evaluation, two prediction target variables were defined to serve as labels for the training and validation as: the aggregated number of requests received for each type of complaint in each month (i.e., complaint count defined as $C_t^i$ in Equation 1), and the complaint ratio, which represents the number of complaints per complaint type divided by total number of complaints received, as shown in Equation 2.

$$C_t^i = \sum_{n=1}^{N} count(i) \qquad (1)$$

$$Y_t^i = \log\left(\frac{C_t^i}{(\sum_{i=1}^{I} C_t^i)/I}\right) \qquad (2)$$

Each type of complaint recorded during the analysis period was grouped monthly to avoid scarcity of data, because some complaints do not occur daily or weekly. In Equation 1, $C_t^i$ is the aggregated count for a given complaint type, $i$, for time $t$, where $i \in \{1, 2, \ldots, 175\}$, and each $i$ represents one type of complaint; $t \in \{1, 2, \ldots, 24\}$, represents the month and year for the complaint, $n \in \{1, 2, \ldots, N\}$ represents the day of month, and $N$ is the total number of days in month $t$. The reason of using $Y_t^i$ (i.e., complaint ratio) as a target variable is to measure importance of a certain type of complaint among all complaint types. The logarithm is used to generate a smooth series of value, in case that the change of complaint ratio is large. For the 27 projects selected in this study, the zone of impact was set to the zip code of each project, and the data analysis period was set to two years.

### 4.3. Selecting Construction Related Complaints from 311 Service Requests

Out of all 175 different types of complaints recorded in the 311 dataset, only a subset were influenced by construction activities. To determine the complaint types that are closely related to the construction projects selected, the t-test was used. The inputs for the t-test were the cumulated complaint count for each complaint type before and after the construction begins. P-values were calculated to find the complaint types that significantly changed with respect to before and after a construction project starts. If the p-value is less than 0.05 (i.e., a 95% confidence level that the distribution of one type of complaint was different before and during the construction), the complaint type was deemed as significant due to the introduction of construction activities, and included in the analysis. Furthermore, to ensure the complaint type was not associated with only a small number of construction projects, only complaint types that occurred more than five times in all 27 construction projects were used to build prediction models. It should be noted that the complaint types determined by the significance test are then filtered based on Table 1, which is the mapping of construction factors to QoL indicators. Complaints that passed the t-test but were not part of Table 1 were also excluded from the analysis. A final list of complaint types selected in this study is shown in Table 3, where a total of 15 complaint types were selected.

**Table 3. Selected Complaint Types in 311 Dataset Related to Construction**

| QoL Indicator | Complaint Types | Frequency |
|---|---|---|
| Environment | Noise | 5 |
| | Noise construction | 5 |
| | Air quality | 5 |
| | Water system, hot/cold water systems, plumbing | 6,7,5 |
| | Street condition | 11 |
| Waste | Solid Waste | 7 |
| | Waste Water (Sewer) | 5 |
| Safety | Project inspection | 6 |
| | Safety | 6 |
| Transportation | Parking | 7 |
| | Metering | 8 |
| Other | Building use | 7 |
| | General construction | 8 |

## 4.4. Prediction Models Built

In this step, we created prediction models to predict the two target variables (i.e., complaint count and complaint ratio). The input for the prediction models were (1) construction related information and (2) complaint count/complaint ratio from before the construction started. Parameter tuning was done to generate the best performing models using three supervised learning algorithms (i.e., Ordinary Least Squares, Random Forest with Adaboost and Decision Trees). For the OLS method, there exists one optimal solution when using Maximum Likelihood algorithm. The result of OLS was used as a baseline comparison with decision trees and random forest. As for RF and DT algorithms, parameter tuning was done by minimizing the MSE value and maximizing the R-Squared value simultaneously. Ten-fold cross validation was executed for each algorithm to get average MSE and R-Squared values for accuracy.

### 4.4.1. Selecting the Best Performing Model for the First Target Variable - Complaint Count

Two parameters in the RF algorithm determine the accuracy, which are depth of tree (i.e., the maximum depth of the base model decision trees) and number of estimators (i.e., the number of different base models created). We tested both parameters in the range of 1 to 20. To create a consistent comparison among various configurations of depth of tree and number of estimators, the input data was normalized to the scale of 0 to 1. The result of RF algorithm with Adaboost for the first target variable (i.e., complaint count $C_t^i$) is shown in Figure 5. From Figure 5a, higher R-Squared value is represented with dark red, whereas dark blue represents lower R-Squared value. Similarly, larger MSE is in dark red, and lower MSE is in dark blue. The highest R-Squared and lowest MSE produced by RF algorithm with Adaboost are highlighted in black squares, when the number of estimators is 12, and depth of tree is 8.

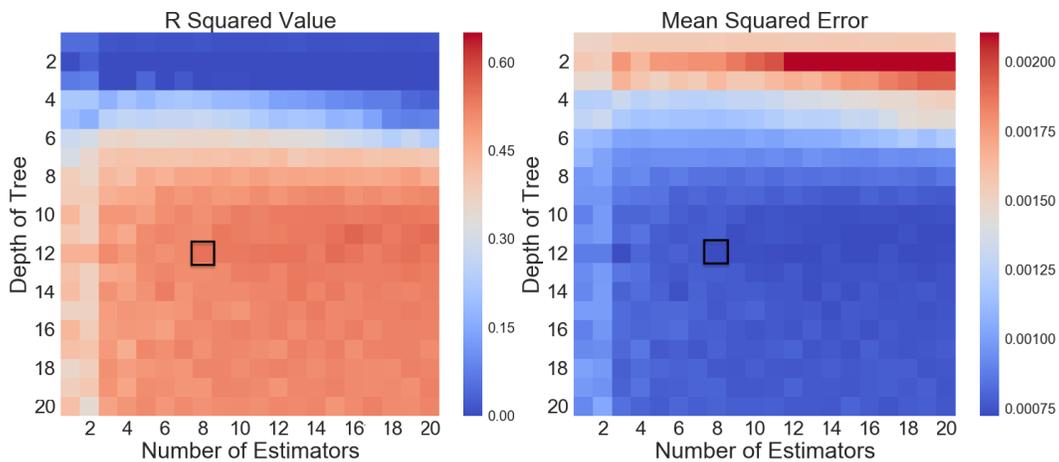

**Figure 5. Results of RF with Adaboost Prediction Model for Complaint Count**.
*Figure 5a (left): R-squared value with changing configurations of the depth of tree (y-axis) and number of estimators. (x-axis)*
*Figure 5b (right): MSE with changing configurations of the depth of tree (y-axis) and number of estimators (x-axis). Larger value is represented with dark red, and smaller value in dark blue.*

Similar to the RF algorithm, DT algorithm's parameter-tuning could also be done by maximizing the R-Squared value and minimizing the MSE. One key difference between DT and RF, is that DT only has one parameter, depth of tree, as a factor to affect the accuracy. The results of DT for the first target variable (i.e., complaint count $C_t^i$) are shown in Figure 6. It is easy to find that, when the depth of tree is set to 8 (as circled in Figure 6), the DT algorithm returns the best results in terms of R-Squared and MSE values.

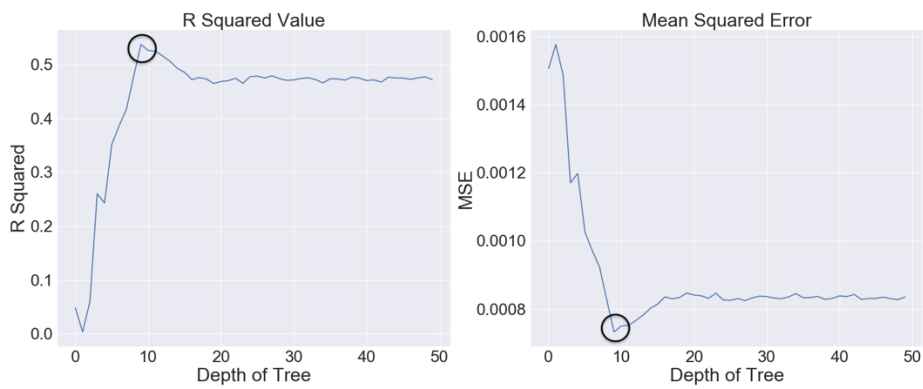

**Figure 6. Results of Decision Tree Prediction Model for Complaint Count.**
*Figure 6a (left): R-Squared value (y-axis) with changing configurations of the depth of tree (x-axis).*
*Figure 6b (right): MSE (y-axis) with changing configurations of the depth of tree (x-axis).*

Finally, the R-Squared values of each model with tuned parameters, are presented in Table 4. It can be seen that the RF algorithm with Adaboost produced the highest R-Squared value among three algorithms. DT with tuned parameters produced a slightly lower R-Squared value than RF, and the OLS algorithm generated the worst performing predictive model with an R-Squared value of 0.20, because of the over-aggressive assumption of data linearity.

**Table 4. R-Squared Value for OLS, DT, and RF with Adaboost for Complaint Count.**

| Algorithm | OLS | DT | RF with Adaboost |
|---|---|---|---|
| R-Squared Value | 0.20 | 0.62 | 0.65 |

### 4.4.2. Selecting the Best Performing Model for the Second Target Variable - Complaint Ratio

Just like the first target variable (i.e., complaint count), the second target variable (i.e., complaint ratio) was also predicted using the three regression algorithms. The parameter tuning and final model selection processes were identical to the ones that were used in the first predicted variable. The results of RF algorithm with Adaboost and DT algorithm are shown in Figure 7 and Figure 8. As seen in Figure 7, the highest R-Squared value and lowest MSE value were achieved when the depth of tree was 12 and number of estimators was 11 for the RF algorithm with Adaboost. For the DT algorithm, as the results in Figure 8 suggest, the DT algorithm achieved the best R-Squared and MSE values when the depth of tree was set to 15.

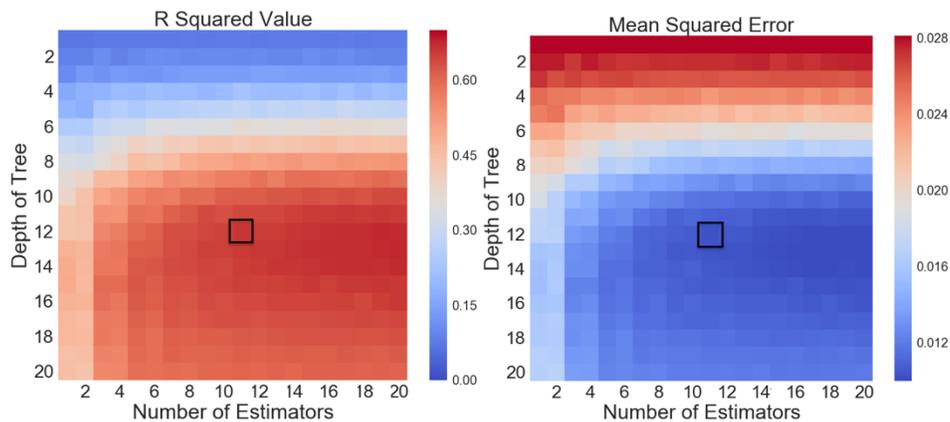

**Figure 7. Results of RF with Adaboost Prediction Model for Complaints Ratio.**

*Figure 7a (left): R-squared value with changing configurations of the depth of tree (y-axis) and number of estimators (x-axis).*

*Figure 7b (right): MSE with changing configurations of the depth of tree (y-axis) and number of estimators (x-axis). Larger value is represented with dark red, and smaller value in dark blue.*

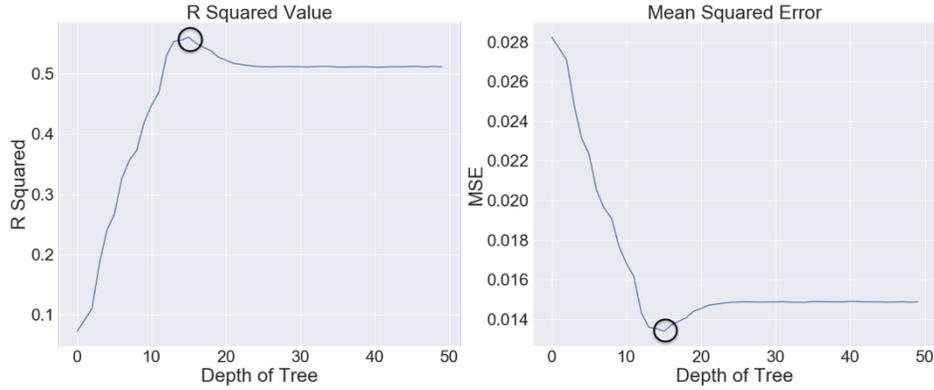

**Figure 8. Results of Decision Tree Prediction Model for Complaints Ratio.**
*Figure 8a (left): R-Squared (y-axis) value with changing configurations of the depth of tree (x-axis).*
*Figure 8b (right): MSE (y-axis) with changing configurations of the depth of tree (x-axis).*

The R-Squared values for the best performing models of complaint ratio are shown in Table 5. It is observed that the RF algorithm with Adaboost again produced the highest R-Squared value, meaning that the model created using RF algorithm performs the best when used for predicting construction complaint ratio defined in Equation 2.

**Table 5. R-Squared Value for OLS, DT, and RF with Adaboost for Complaint Ratio.**

| Algorithm | OLS | DT | RF with Adaboost |
|---|---|---|---|
| R-Squared Value | 0.17 | 0.58 | 0.67 |

### 4.5. Predicting Construction related Complaints Counts

During the ten-fold cross validation for each algorithm, we monitored the predicted complaint count related to construction before and after the construction started, and compared the predicted complaint count with the actual complaint counts received during construction. For the best performing model (i.e., random forest), we show, in Table 6, the average predicted complaint count changes vs. the actual complaint count changes for a subset of complaints. Results from Table 6 show that our prediction models successfully predicted the increase of various complaint types regarding construction activities, though the exact percentage values were not strictly the same as the actual percentage change.

**Table 6. Percentage Change for Complaint Counts for a Set of Complaint Types**

| Complaint Type | Predicted Change in Complaint Count* | Actual Change in Complaint Count* |
|---|---|---|
| Air Quality | 71% (month 1-4)* | 55% (month 1-4)* |
| Sewer | 23% (month 1-3) | 12% (month 1-3) |
| Safety | 52% (month 4-8) | 31% (month 4-8) |
| Noise Construction | 10% (month 1-3) | 19% (month 1-3) |
| Unsanitary Street Condition | 7% (month 10-12) | 3% (month 10-12) |

*Information in the parenthesis is showing the time-frame that the change happened for the corresponding complaint type with respect to the start of the construction project.

These detailed results show correlations with schedule of activities at the beginning (e.g., excavation, earthwork, underground site condition improvements resulting in air quality, noise and sewer issues around) and the later stage of construction projects (e.g., construction waste such as dust blown out of construction sites, and safety issues caused by heavy machinery) and provide insights for construction companies and government agencies to understand how construction projects are affecting surrounding residents' QoL in certain periods of construction.

In this case, the residents were experiencing dirty street conditions at the end of construction, and extensive noise and low air quality at the beginning of the construction. The increase in the noise and air quality complaints toward the start of construction could be explained by the heavy machinery used during the excavation period for road reconstruction projects. The increase in sewer related complaints are caused by the excavation, and interrupting the sewerage system when working around/on pipes. The results show that, construction companies can potentially use this information to actively prevent noise pollution at the beginning, inform public about possible issues with water and sewer systems, and put more efforts into cleaning the environment towards the end of construction.

## 5. CONCLUSION AND FUTURE WORK

This paper proposed a machine learning based approach to quantify and predict the impact of construction projects on urban QoL. The approach was implemented on 27 road reconstruction projects managed by NYC DDC in the borough of Manhattan. Three supervised learning algorithms, Ordinary Least Squares, Decision Trees and Random Forest with Adaboost, were used for building prediction models. Two target variables, the complaint count and complaint ratio, were defined for prediction accuracy test. Results show that Random Forest with Adaboost produced the best performing prediction model for both target variables. The prediction results show that the final prediction model achieved R-Squared values of 0.65 and 0.67 on both target variables. Empirical implications of the prediction results were discussed (e.g., air quality, construction noise and sewage related complaints increased as soon as the construction starts; safety and street sanitary related complaints increased in the later part of the construction period). It is shown that the approach could potentially provide insights for construction companies and government agencies to understand how construction projects are affecting surrounding residents' QoL. Practitioners can use the generic steps provided for this approach to build models using the data available from their localities to quantify the impact of new construction projects on certain QoL indicators.

Future work is needed to improve the current research. This paper only tested the approach using limited construction projects in a specific city as testbeds. Extended studies of the approach on other types of construction projects and cities will be needed. This work will also be extended to include other types of open city data that can complement the 311 dataset, such as emergency response data, through which impact of construction on traffic conditions can better be quantified.

In conclusion, this paper is a first attempt on using open city data and construction related data to quantify and predict the impact of construction projects on urban QoL. This study could serve as a guide for future construction related QoL research, in a sense that it utilized open data and machine learning algorithms instead of traditional survey studies. This paper also provided a fresh angle for looking at the impact of construction projects by integrating open city data sets and data driven approaches.

## DATA AVAILABILITY

- Some or all data, models, or code used during the study were provided by a third party (NYC Open Data). Direct access of the data can be found at NYC open data portal.